\documentclass[showpacs,indentfirst,nofootinbib,12pt]{revtex4}
\usepackage{epsfig}
\usepackage{epsf}
\usepackage{amsmath}
\usepackage{color}
\usepackage{bm}
\usepackage{multirow}
\begin{document}

\title{Hadronic-loop induced mass shifts in scalar heavy-light mesons}

\author{Feng-Kun Guo$^1$}
\email{f.k.guo@fz-juelich.de}%
\author{Siegfried Krewald$^1$}
\author{Ulf-G. Mei\ss ner$^{1,2}$}
\affiliation{\small $^1$Institut f\"ur Kernphysik (Theorie),
Forschungszentrum J\"ulich, D-52425 J\"ulich, Germany\\
$^2$Helmholtz-Institut f\"ur Strahlen- und Kernphysik (Theorie)
Universit\"at Bonn, Nu{\ss}allee 14-16, D-53115 Bonn, Germany}

\date{\today}

\begin{abstract}
We calculate the mass shifts of heavy-light scalar mesons due to
hadronic loops under the assumption that these vanish for the
groundstate heavy-light mesons.  The results show that the masses
calculated in quark models can be reduced significantly. We stress
that the mass alone is not a signal for a molecular interpretation.
Both the resulting mass and the width suggest the observed $D_0^*$
state could be a dressed $c\bar q$ state. We give further
predictions for the bottom scalar mesons which can be used to test
the dressing mechanism.
\end{abstract}

%\keywords{}

\maketitle

\section{Introduction}

\begin{figure}[htbp]
\begin{center}\vspace*{0.0cm}
\includegraphics[width=0.6\textwidth]{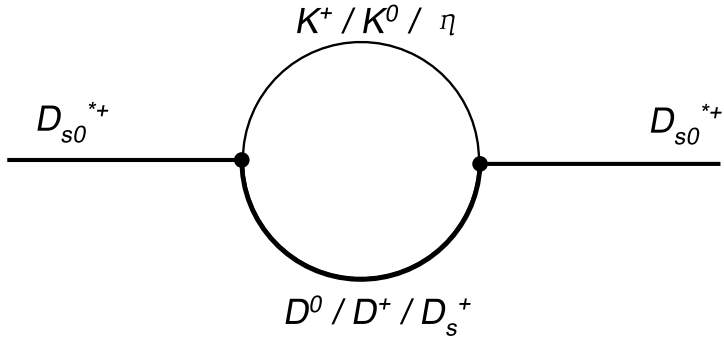}%
\vglue -0.5cm\caption{\label{fig:loop}The relevant hadronic loops
coupled to the $D_{s0}^*$.}
\end{center}
\end{figure}

The constituent quark model has been very successful in describing
hadron spectroscopy. In recent years, some newly observed hadrons
attracted much interest from both the experimental and theoretical
community since these hadrons do not fit to the quark model
predictions \cite{Swanson:2006st}. For instance, the mass of the
observed charm-strange scalar meson $D_{s0}^*$ \cite{Aubert:2003fg}
is $2317.3\pm0.6$ MeV \cite{Yao:2006px}, while the predictions from
most quark models spread from 2400~MeV to 2500~MeV
\cite{godfrey,qm2317}. The high mass predicted in constituent quark
models is obtained through an orbital angular momentum excitation.
The result from QCD sum rules in heavy quark effective theory gives
a mass range $2.42\pm0.13$~GeV for the $D_{s0}^*$ ($c\bar s$) state
which is consistent with, but the central value is 100 MeV higher
than, the experimental value~\cite{Dai:2003yg}.
Due to the fact that the mass of the $D_{s0}^{*}(2317)$ is just
below the $DK$ threshold at 2.36~GeV, a $DK$ molecular
interpretation was proposed by Barnes et al.~\cite{Barnes:2003dj}
and some others~\cite{DK}. We want to remark that a $DK$ bound state
can be dynamically generated with a mass consistent with the
observed mass of the $D_{s0}^{*}(2317)$ in the framework of the
heavy chiral unitary approach~\cite{Hchua,Guo:2006fu}. Other exotic
explanations were also proposed, such as tetraquark
state~\cite{tetraq}, and $D\pi$ atom~\cite{Dspi}.
Besides these  exotic explanations, some authors tried to modify the
quark model predictions. In~\cite{Lakhina:2006fy}, one loop
corrections to the spin-dependent one-gluon exchange potential was
considered, and the predicted mass of the $D_{s0}^*$ is higher than
the experimental value by only about 20 MeV. Another kind of
modification is the mixing of the $c{\bar s}$ with the $cq{\bar
s}{\bar q}$ tetraquark~\cite{Browder:2003fk}, or considering the
coupling of the $c{\bar s}$ to hadronic channels, such as
$DK$~\cite{van Beveren:2003kd}. The prediction for the $D_{s0}^*$
from the QCD sum rules can also be lowered to $2.331\pm0.016$~GeV
considering the $DK$ continuum explicitly~\cite{Dai:2006uz}. Note
that it is possible to distinguish a hadronic bound state from
elementary hadrons, as done for the deuteron~\cite{Weinberg:1965}
and for the light scalars $a_0(980)$ and
$f_0(980)$~\cite{Baru:2003qq}. All the results from Refs.~\cite{van
Beveren:2003kd,Dai:2006uz} indicate the importance of the strongly
coupled hadronic channels on determining the mass of a hadron.
For the charm-non-strange sector, both the Belle and FOCUS
collaborations reported a scalar meson with a large
width~\cite{Abe:2003zm,Link:2003bd}. Although the reported masses by
different collaborations are not consistent with each other, the
measurements are considered as the same charm scalar meson by the
Particle Data Group~(PDG), and the PDG average value of the mass is
$2352\pm50$~MeV. The structure of this state has not been clear yet.
In this Letter, we revisit the mass shifts of the heavy scalar
mesons induced by the strongly coupled hadronic loops. For instance,
the $D_{s0}^{*+}$, the $1^3P_0$ $c\bar s$ state in quark model, can
couple to the $D^+K^0$, $D^0K^+$ and $D_s\eta$ loops, see
Fig.~\ref{fig:loop}. The coupling constants of the $D_{s0}^{*+}$ to
the three channels can be related by SU(3) symmetry. We shall use
three different coupling types to study the mass shifts, called
Model I, II and III in the following. In Model I, the coupling of
the scalar heavy meson~($S$) to the heavy pseudoscalar meson~($P$)
and the Goldstone boson~($\phi$) is assumed to be a constant. In
Model II, the coupling is derived in the framework of heavy meson
chiral perturbation theory (HM$\chi$PT) which combines the chiral
expansion with the heavy quark expansion
\cite{Wise:1992hn,Kilian:1992hq} (for a review, see
Ref.~\cite{Casalbuoni:1996pg}). In Model III, a chiral effective
coupling is constructed disregarding the heavy quark expansion. Of
course, such corrections due to hadronic loops are always
model-dependent and are, in general, difficult to quantify, for a
recent discussion see \cite{Capstick:2007tv} (For further discussion
of incorporating hadronic loops in the quark model,
see~\cite{Barnes:2007xu}). This is why we consider three different
models and also need to assume that for the ground state meson
$Q\bar q$ (with $Q=c,b$ and $q = u,d,s$), the shift due to the
hadronic loops vanishes as it was done e.g. in the calculation of
the mass shifts of charmonia in Ref.~\cite{Pennington:2007xr}.

%--------------------------------------------------------------------------------
%--------------------------------------------------------------------------------

\section{Choices of coupling}
\subsection{Model I}
First, the coupling of $S$ to $P,\phi$ is taken as a constant.
The loop that integral enters the dressed propagator is
\begin{equation}
G^{\rm I}(s) = i\int {d^4q\over (2\pi)^4} {1\over
(q^2-m_1^2+i\epsilon)[(p-q)^2-m_2^2+i\epsilon]},
\end{equation}
where $s=p^2$. The analytic expression is given
by~\cite{Oller:1998zr,Oller:2000fj}
\begin{eqnarray}
\label{eq:GI} G^{\rm I}(s) &\!=&\!
\frac{1}{16\pi^2}\left\{R-1+\ln{\frac{m_2^2}{\mu^2}} +
\frac{m_1^2-m_2^2+s}{2s}\ln{\frac{m_1^2}{m_2^2}} +\frac{\sigma}{2s}\left[\ln({s-m_1^2+m_2^2+\sigma})\right.\right.\nonumber\\
&\!&\! \left.\left. -\ln({-s+m_1^2-m_2^2+\sigma}) +
\ln({s+m_1^2-m_2^2+\sigma})-\ln({-s-m_1^2+m_2^2+\sigma}) \right]
\right\},
\end{eqnarray}
where $R=-[2/(4-d)-\gamma_E+\ln(4\pi)+1]$ will be set to zero in the
calculations and $\gamma_E$ is Euler's constant, $\mu$ is the scale
of dimensional regularization, and
$\sigma=\sqrt{[s-(m_1+m_2)^2][s-(m_1-m_2)^2]}$.
Taking into account the $D^0K^+$, $D^+K^0$ and $D_s^+\eta$ channels,
the shifted mass of the $D_{s0}^{*+}$ is given by the solution of
the equation
\begin{equation}
s - (\overset{\circ}M_{D_{s0}^*})^2 - g^2 {\rm Re}\left[2G^{\rm
I}_{DK}(s)+{2\over3}G^{\rm I}_{D_s\eta}(s) \right]= 0,
\end{equation}
where $\overset{\circ}M_{D_{s0}^*}$ denotes the bare mass of the
$D_{s0}^*$.
%-------------------
Note that we only consider the lowest possible intermediate states.
In principle, all states with quantum numbers allowed by
conservation laws can contribute~\cite{Barnes:2007xu}. For instance,
besides the channels considered here, the $D^*K^*$, $D_s\eta'$ and
$D_s^*\rho$ can contribute either. But their threshold are at least
500~MeV higher than that of the DK, thus their contributions are
expected to be suppressed.
%-------------------
The corresponding equation for the non-strange charm meson
$D_0^{*+}$ is
\begin{equation}
s - (\overset{\circ}M_{D_{0}^*})^2 - g^2 {\rm
Re}\left[{3\over2}G^{\rm I}_{D\pi}(s)+{1\over6}G^{\rm I}_{D\eta}(s)
+G^{\rm I}_{D_sK}(s) \right]= 0,
\end{equation}
where four channels, $D^+\pi^0$, $D^0\pi^+$, $D^+\eta$ and
$D_s^+K^0$, are taken into account.
The coupling constant $g$ has been calculated by using light-cone
QCD sum rules in \cite{Colangelo:1995ph}, $g=6.3\pm1.2$~GeV in the
charm sector, $g=21\pm7$~GeV in the bottom sector. A more recent
analysis considering the $D_{s0}^*(2317)$ state as a conventional
$c{\bar s}$ meson gives the coupling constant for $D_{s0}^*DK$ as
$5.9^{+1.7}_{-1.6}$ GeV \cite{Wang:2006ida}, which is consistent
with that given in Ref.~\cite{Colangelo:1995ph}. In this Letter, we
study the mass shifts of bare $c{\bar q}$ (and $b{\bar q}$) mesons
induced by hadronic loops. The values of coupling constants given in
Ref.~\cite{Colangelo:1995ph} will be taken because the masses of the
scalar heavy mesons used therein are consistent with the quark-model
expectation (no fitting to the mass of the $D_{s0}^*(2317)$ was
performed since the state had not been discovered yet) and hence
correspond to the bare masses.

\subsection{Model II}

In the HM$\chi$PT, the Lagrangian for the coupling of $S$ to $P$ and
$\phi$ to leading order is~\cite{Kilian:1992hq,Casalbuoni:1996pg}
\begin{eqnarray}
{\cal L} &\!=&\! ih\langle S_b\gamma_{\mu}\gamma_5A^{\mu}_{ba}{\bar
H_a}\rangle + h.c. \nonumber\\
&\!=&\! {i\sqrt{2}h\over f_{\pi}} \left(
D_{0b}v^{\mu}\partial_{\mu}\Phi_{ba} P^{\dag}_a -
D_{1b}^{\nu}v^{\mu}\partial_{\mu}\Phi_{ba} P^{*\dag}_{a\nu} \right)
+ \cdots,
\end{eqnarray}
where the axial field
\begin{equation}
A^{\mu}_{ba} = {i\over2} (\xi^\dag \partial^\mu \xi-\xi\partial^\mu
\xi^\dag)_{ba} = - {\partial^\mu\Phi_{ba}\over\sqrt{2}f_{\pi}} +
\cdots
\end{equation}
contains the Goldstone bosons
\begin{eqnarray}
\label{eq:ps} \Phi = \frac{1}{\sqrt{2}}\sum_{a=1}^8 \lambda_a \phi_a
= \left(\begin{array}{ccc}
\frac{1}{\sqrt{2}}\pi^0+\frac{1}{\sqrt{6}}\eta & \pi^+ & K^+\\
\pi^- & -\frac{1}{\sqrt{2}}\pi^0+\frac{1}{\sqrt{6}}\eta & K^0\\
K^- & \bar{K}^0 & -\frac{1}{\sqrt{3}}\eta
\end{array}\right),
\end{eqnarray}
the subscripts $a,b$ represent to the light quark flavor $u,d,s$,
and $f_{\pi}=92.4$ MeV is the pion decay constant.
\begin{equation}
\label{eq:H} H_a = {1+\not\!
v\over2}\left(P^{*\mu}_a\gamma_{\mu}-P_a\gamma_5\right)
\end{equation}
represents the multiplet containing the pseudoscalar charm mesons,
$P=(D^0,D^+,D_s^+)$, and vector charm mesons,
$P^*=(D^{*0},D^{*+},D_s^{*+})$, and ${\bar
H}=\gamma_0H^\dag\gamma_0$.
\begin{equation}
\label{eq:S} S_a = {1+\not\!
v\over2}\left(D^{\mu}_{1a}\gamma_{\mu}\gamma_5-D_{0a}\right),
\end{equation}
represents the multiplet containing the scalar charm mesons,
$D_0=(D^{*0}_0,D^{*+}_0,D_{s0}^{*+})$, and axial charm mesons,
$D_1=(D_1^0,D_1^+,D_{s1}^+)$. These field operators in Eqs.
(\ref{eq:H}) and (\ref{eq:S}) have dimension 3/2 since they contain
a factor $\sqrt{M}$, where $M$ is the mass of the corresponding
meson, in their definition.
Let $p$ denotes the momentum of a pseudoscalar charm meson, e.g.
$D$, and $k=p-M_Dv$ its residual momentum. The propagator of $D$ in
HM$\chi$PT is~\cite{Wise:1992hn}
\begin{equation}
{i\over 2(v\cdot k + \frac{3}{4}\Delta)},
\end{equation}
where $\Delta=M_{D^*}-M_D$. The propagator of the strange charm
meson $D_s$ is
\begin{equation}
{i\over 2(v\cdot k + \frac{3}{4}\Delta_s-\delta)},
\end{equation}
where $\Delta_s=M_{D^*}-M_D$ and $\delta=M_{D_s}-M_D$. In the actual
calculations, we take $\delta=0.1$~GeV which is an approximate value
of $M_{D_s}-M_D$ and $M_{D_s^*}-M_{D^*}$. The propagator of the
scalar charm mesons are similar with proper mass differences, i.e.
$\Delta_S=M_{D_1}-M_{D_0^*}$ for $D_0^*$,
$\Delta_{Ss}=M_{D_{s1}}-M_{D_{s0}^*}$ for $D_{s0}^*$, and the SU(3)
breaking mass difference can be taken as $\delta_S=0.1$ GeV, the
same as $\delta$.
The coupling of the $D_0^{*+}$ to the $D^0$ and $\pi^+$ to leading
order is
\begin{eqnarray}
i\langle\pi^+(q)D^0(q')|{\cal L}|D_0^{*+}(p)\rangle &\!=&\!
-i{\sqrt{2}h\over f_{\pi}}\sqrt{M_D\overset{\circ}M_{D_0^*}}v\cdot q \nonumber\\
&\!=&\! -i{h\over \sqrt{2}f_{\pi}}\sqrt{M_D\overset{\circ}M_{D_0^*}}
{(\overset{\circ}M_{D_0^*})^2-M_D^2+m_{\pi}^2\over
\overset{\circ}M_{D_0^*}}.
\end{eqnarray}
The last equality holds for on-shell $D^0$ and $\pi^+$ mesons. The relation
between $h$ and $g$ is
\begin{eqnarray}
h = - {\sqrt{2}f_{\pi}\overset{\circ}M_{D_0^*}\over
\sqrt{M_D\overset{\circ}M_{D_0^*}}
\left((\overset{\circ}M_{D_0^*})^2-M_D^2+m_{\pi}^2\right)} g.
\end{eqnarray}
Corresponding to $g=6.3\pm 1.2$ GeV, calculations via QCD sum rules
give $h=-0.44\pm0.09$ for charm mesons~\cite{Colangelo:1995ph}. In
Ref.~\cite{Colangelo:1995ph}, the mass difference between the scalar
and the pseudoscalar heavy meson is taken to be 500~MeV. For bottom
mesons, we have $h=-0.52\pm0.18$~\cite{Colangelo:1995ph}.
The momentum of a scalar charm meson is $p=\overset{\circ}M_{S}v+k$,
where $\overset{\circ}M_{S}$ is the bare mass of the scalar charm
meson and $k$ is the residual momentum. The momentum of the
Goldstone boson in the loop is denoted by $q$. Then the residual
momentum of the pseudoscalar charm meson, whose mass is $M_P$, in
the loop should be $k'=p-q-M_Pv=k+(\overset{\circ}M_{S}-M_P)v-q$.
The loop integral is
\begin{equation}
G^{\rm II}(v\cdot k) = {i\over2}\int {d^4q\over (2\pi)^4} {(v\cdot
q)^2\over
(q^2-m^2+i\varepsilon)[(v\cdot(k-q)+\overset{\circ}M_S-M_P+\Delta+i\varepsilon]},
\end{equation}
where $m$ is the mass of the Goldstone boson in the loop. The loop
integral can be worked out as~\cite{Bernard:1995dp,Scherer:2002tk}
\begin{eqnarray}
\label{eq:GII} G^{\rm II}(v\cdot k) = {m\over 16\pi^2} J(0;\omega)
\end{eqnarray}
and
\begin{eqnarray}
J(0;\omega) &\!=&\! \omega(R+\ln{m^2\over \mu^2}-1) \nonumber\\
&\!&\!+ \begin{cases}
      2\sqrt{\omega^2-m^2}\cosh^{-1}({\omega\over m}) - 2\pi
      i\sqrt{\omega^2-m^2}, & \omega>m \\
      2\sqrt{m^2-\omega^2}\cos^{-1}(-{\omega\over m}), &
      \omega^2<m^2 \\
      -2\sqrt{\omega^2-m^2}\cosh^{-1}(-{\omega\over m}), & \omega<-m
      \end{cases}
\end{eqnarray}
where $\omega=v\cdot k+\overset{\circ}M_S-M_P+3\Delta/4$.
The dressed propagator for $D_{s0}^{*+}$ is
\begin{equation}
{i \over 2\left(v\cdot k + {3\over4}\Delta_{Ss}-\delta_S\right) -
{2h^2\over f_{\pi}^2}\overset{\circ}M_{D_{s0}^*} {\rm
Re}\left[2M_DG^{\rm II}_{DK}(v\cdot k)+{2\over3}M_{D_s}G^{\rm
II}_{D_s\eta}(v\cdot k) \right]}.
\end{equation}
For the mass difference $\Delta_{Ss}$ the physical values is taken
in the propagator which is correct to the order considered. Thus
substituting $v\cdot k$ by $v\cdot p-\overset{\circ}M_{D_{s0}^*}$,
the shifted mass can be given the value of $v\cdot p$ which is the
solution of the following equation \cite{Bernard:1995dp}
\begin{equation}
 2\left(v\cdot p-\overset{\circ}M_{D_{s0}^*}\right) -
{2h^2\over f_{\pi}^2}\overset{\circ}M_{D_{s0}^*} {\rm
Re}\left[2M_DG^{\rm II}_{DK}(v\cdot k)+{2\over3}M_{D_s}G^{\rm
II}_{D_s\eta}(v\cdot k) \right]=0.
\end{equation}
The corresponding propagator of the $D_0^*$ is
\begin{equation}
{i \over 2\left(v\cdot k + {3\over4}\Delta_{S}\right) - {2h^2\over
f_{\pi}^2}\overset{\circ}M_{D_{0}^*} {\rm
Re}\left[{3\over2}M_DG^{\rm II}_{D\pi}(v\cdot k)+{1\over6}M_DG^{\rm
II}_{D\eta}(v\cdot k) +M_{D_s}G^{\rm II}_{D_sK}(v\cdot k) \right]}.
\end{equation}
From Eq.~(\ref{eq:GII}), the mass shift vanishes in the chiral
limit, similar to the mass shift of the nucleon due to $N\pi$ loop
in heavy baryon $\chi$PT \cite{Bernard:1995dp}, in contrast to that
in Model I and III, see below.

\subsection{Model III}

Disregarding the heavy quark expansion, we can directly construct an
effective chiral Lagrangian describing the coupling of a scalar
charm meson with a pseudoscalar charm meson and a Goldstone boson.
The Lagrangian is
\begin{eqnarray}
\label{eq:model3} {\cal L} =
h'D_{0b}A^{\mu}_{ba}\partial_{\mu}P^\dag_a+ {\rm h.c}\, .
\end{eqnarray}
Note that the field operators of the heavy mesons in Eq.~(\ref{eq:model3})
have dimension 1, different from those in Eqs.~(\ref{eq:H},\ref{eq:S}).
This Lagrangian drives the coupling to be of the type
\begin{eqnarray}
i\langle\pi^+(q)D^0(q')|{\cal L}|D_0^{*+}(p)\rangle &\!=&\!
i{h'\over \sqrt{2}f_{\pi}}q\cdot q' \nonumber\\
&\!=&\! i{h'\over 2\sqrt{2}f_{\pi}}
\left((\overset{\circ}M_{D_0^*})^2-M_D^2-m_{\pi}^2\right).
\end{eqnarray}
The second equality is fulfilled only for on-shell $D^0$ and
$\pi^+$ mesons. The relation between $h'$ and $g$ is
\begin{eqnarray}
h'= {2\sqrt{2}f_{\pi}\over
(\overset{\circ}M_{D_0^*})^2-M_D^2-m_{\pi}^2}g.
\end{eqnarray}
For extracting the value of $h'$, we take
$\overset{\circ}M_{D_0^*}-M_D=500$ MeV following
Ref.~\cite{Colangelo:1995ph}.
Then we have $h'=0.78\pm0.15$ for charm
mesons. Similarly, the coupling constant for bottom mesons can be
obtained as $h'=1.00\pm0.33$.
The dressed propagators of the $D_{s0}^{*+}$ and $D_0^{*+}$ are
\begin{equation}
{i \over s - (\overset{\circ}M_{D_{s0}^*})^2 - {h'^2\over2f_{\pi}^2}
{\rm Re}\left[2G^{\rm III}_{DK}(s)+{2\over3}G^{\rm III}_{D_s\eta}(s)
\right]},
\end{equation}
and
\begin{equation}
{i \over s - (\overset{\circ}M_{D_0^*})^2 - {h'^2\over2f_{\pi}^2}
{\rm Re}\left[{3\over2}G^{\rm III}_{D\pi}(s)+{1\over6}G^{\rm
III}_{D\eta}(s) +G^{\rm III}_{D_sK}(s) \right]},
\end{equation}
respectively. The physical masses of the charm scalar mesons can be
obtained by setting the denominators of the propagators to zero.
The loop integral in the dressed propagators is
\begin{equation}
G^{\rm III}(s) = i\int {d^4q\over (2\pi)^4} {[q\cdot(p-q)]^2\over
(q^2-m_1^2+i\epsilon)[(p-q)^2-m_2^2+i\epsilon]}.
\end{equation}
The analytic expression can be worked out as
\begin{eqnarray}
\label{eq:GIII} G^{\rm III}(s) &\!=&\! \frac{1}{16\pi^2}\left\{
\left[m_1^4+m_1^2m_2^2+m_2^4-{3\over4}(m_1^2+m_2^2)s
+{s^2\over4}\right]\left(R+\ln{\frac{m_2^2}{\mu^2}}\right)
\right.\nonumber\\ &\!&\!\left. -{(m_1^2+m_2^2-s)^2\over4} +
{(m_1^2-m_2^2+s)\left[\sigma^2+2m_1^2(m_1^2+m_2^2)\right]-2m_1^2\sigma^2\over
8s} \ln{\frac{m_1^2}{m_2^2}} \right.\nonumber\\
&\!&\! \left. +
\frac{\sigma(m_1^2+m_2^2-s)^2}{8s}\left[\ln({s-m_1^2+m_2^2+\sigma})
-\ln({-s+m_1^2-m_2^2+\sigma})\right.\right.\nonumber\\
&\!&\! \left.\left. +
\ln({s+m_1^2-m_2^2+\sigma})-\ln({-s-m_1^2+m_2^2+\sigma}) \right]
\right\}~.
\end{eqnarray}

\section{Results}

\begin{figure}[htbp]
\begin{center}\vspace*{0.0cm}
\includegraphics[width=0.48\textwidth]{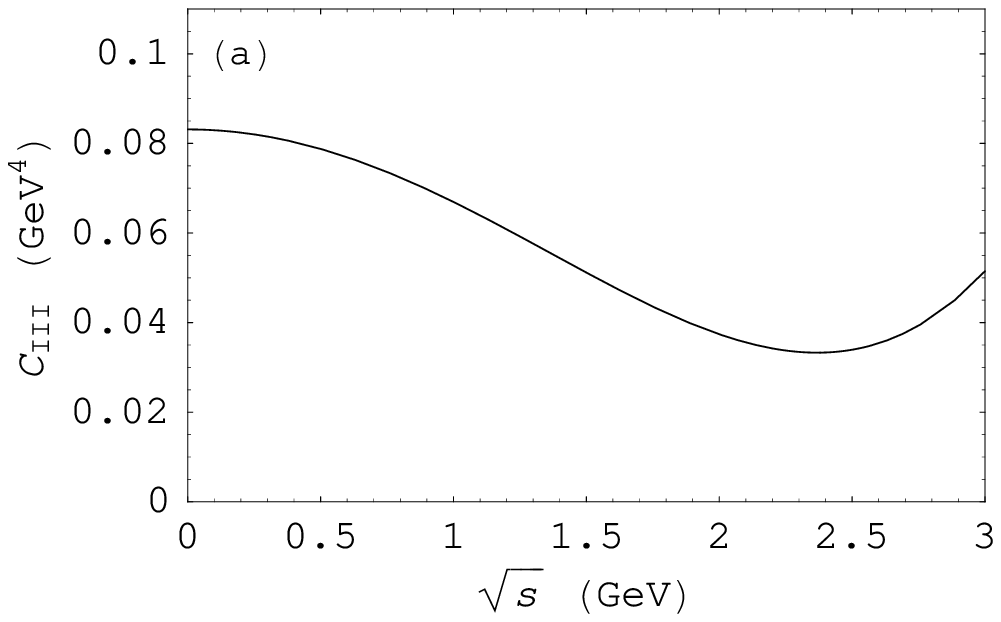}%
\includegraphics[width=0.50\textwidth]{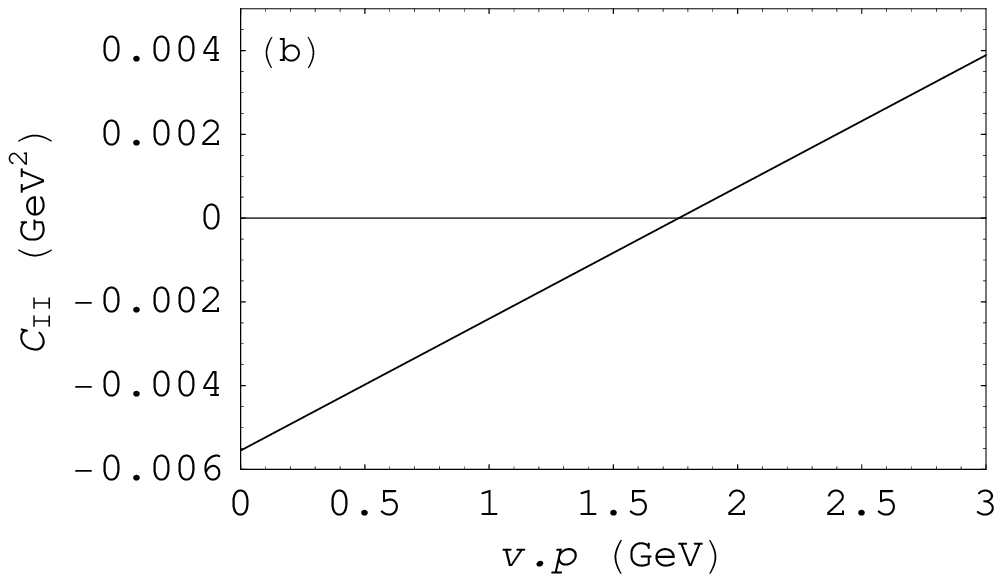}%
\vglue -0.5cm\caption{\label{fig:Coe}The coefficients of the
logarithm $\ln{({m_2^2}/{\mu^2})}$ in the loop functions in Model
III (a) and Model~II~(b). $m_1=M_D$ and $m_2=m_K$ are taken.}
\end{center}
\end{figure}

In general, the effect of a hadronic loop coupling to a bare state
is to pull its bare mass down to the physical one (if the physical
mass is above the threshold of the channel, the imaginary part of
the loop will contribute to the width of the state). One can expect
that the mass of a charm meson cannot be pulled down to below the
mass of the charm quark $m_c$. So we can assume that at some point,
called the subtraction point, above $m_c$, the contribution from any
hadronic loop vanishes. On the other hand, the subtraction point can
not be too high since a state close to the threshold of a strong
decay channel would have a strong coupling to the channel, and hence
its mass would be affected. A similar idea has been taken to study
the mass shifts of charmonia \cite{Pennington:2007xr}.
After choosing a specific subtraction point, the renormalization
scale $\mu$ considered as a parameter can be determined from this
assumption. In the loop function in Model I, Eq.~(\ref{eq:GI}), the
coefficient of the chiral logarithm $\ln{({m_2^2}/{\mu^2)}}$ is a
constant $1/(16\pi^2)$; in the loop functions in Model II and III,
Eq.~(\ref{eq:GII}) and Eq.~(\ref{eq:GIII}), the coefficients of the
same logarithm are momentum-dependent. In Model III, the coefficient
$$C_{\rm III}=\frac{1}{16\pi^2}
\left[m_1^4+m_1^2m_2^2+m_2^4-{3\over4}(m_1^2+m_2^2)s+{1\over4}s^2\right]$$
is always positive, and it changes slowly with respect to $\sqrt{s}$
below 3~GeV for the $DK$ loop, see Fig.~\ref{fig:Coe}(a). So in
Model I and III, the dependence of $\mu$ on the choice of
subtraction point is small. For instance, the values of $\mu$
determined when the subtraction point is chosen at
$\sqrt{s}=M_D=1.87$~GeV and $\sqrt{s}=m_c=1.35$~GeV are listed in
Table~\ref{tab:mu}. One can see when the subtraction point is
changed from $M_D$ to $m_c$, the resulting values of $\mu$ in the
$DK$ loop changes slightly, and all the values are not far away from
$M_D$. In the following, we shall take $M_D$ as the subtraction
point for Model I and III. However, in Model II, the coefficient
$C_{\rm II}=m^2\omega/(16\pi^2)$ is proportional to $v\cdot p$, and
changes its sign at just below $M_D$. In Fig.~\ref{fig:Coe}(b), we
show $C_{\rm II}$ as a function of $v\cdot p$ where $M_P=M_D$ and
$m=m_K$ are used. So the resulting value of $\mu$ depends strongly
on the choice of subtraction point. The values of $\mu$ in Model II
are also given in Table~\ref{tab:mu}. It seems that these values are
unphysical. Because we only take the leading order in heavy quark
expansion to describe the coupling in Model II, the unphysical value
of $\mu$ might indicate that higher order contributions, which will
change the behavior of $C_{\rm II}$, are important. To avoid this
problem, a commonly used value $\mu=1$~GeV in HM$\chi$PT is taken in
Model II. Certainly, the $\mu$ dependence should be absorbed by
counterterms at the next order.
%--------------------------------------------------------------------------------
%--------------------------------------------------------------------------------
%-------------------------------------------------------------------
\begin{table}[htb]
\begin{center}
\caption{\label{tab:mu}The values of the renormalization scale $\mu$
in the $DK$ loop determined when the subtraction point being chosen
at $M_D$ and $m_c$. All units are in GeV.}
\begin{ruledtabular}
\begin{tabular}{lccc}
Subtraction point & Model I & {Model II} & {Model III} \\\hline%
$M_D=1.87$ GeV & 1.58 & 1095.3 & 1.85 \\%
$m_c=1.35$ GeV & 1.84 & 0.20 & 1.82 \\
\end{tabular}
\end{ruledtabular}
\end{center}
\end{table}
%-------------------------------------------------------------------
%
The same method can be used directly for the scalar bottom mesons
with no more free parameters. In Model I and III, we use the same
assumption that the contribution from any hadronic loop vanish at
the mass of the lowest heavy flavor meson. That is, in the bottom
case the subtraction point is chosen to be $M_B=5.28$~GeV. The bare
masses of the scalar heavy mesons are taken from the popular
Godfrey-Isgur quark model which can describe the meson spectroscopy,
especially the low-lying states very well \cite{godfrey},%---------------------------
\footnote{If the parameters of a quark model, e.g. the constituent
quark masses and strong coupling constant, were determined by
fitting to the whole hadron mass spectrum, loop effects would be
incorporated to some extent although in an unclear way. But in
reality, the quark model parameters were fitted to mostly the low
lying states. Especially in the Godfrey-Isgur quark model the
physical spectrum used in fit does not contain the heavy scalar
mesons studied here. }
%---------------------------
i.e. $\overset{\circ}M_{D_{s0}^*}=2.48$ GeV,
$\overset{\circ}M_{D_{0}^*}=2.40$ GeV,
$\overset{\circ}M_{B_{s0}^*}=5.83$ GeV and
$\overset{\circ}M_{B_{0}^*}=5.76$ GeV.

The resulting masses of all the four mesons are listed in
Table~\ref{tab:results}.
%-------------------------------------------------------------------
\begin{table}[htb]
\begin{center}
\caption{\label{tab:results}The resulted masses of the lowest scalar
heavy mesons from dressing in different models.}
\begin{ruledtabular}
\begin{tabular}{lccccc}
 & Bare Mass & Model I & \multicolumn{1}{c}{Model II} & Model III & $\Delta M_{\rm III}$ (GeV) \\\hline%
$M_{D_{s0}^*}$ (GeV) & 2.48 & 2.33-2.39 & 2.26-2.35 & 2.36-2.40 & 0.08-0.12 \\%
$M_{D_{0}^*}$ (GeV) & 2.40 & 2.30-2.35 & 2.39-2.40 & 2.30-2.34 & 0.06-0.10 \\
$M_{B_{s0}^*}$ (GeV) & 5.83 & 5.58-5.72 & 5.73-5.73 & 5.62-5.70 & 0.13-0.21 \\%
$M_{B_{0}^*}$ (GeV) & 5.76 & 5.49-5.67 & 5.85-5.86 & 5.55-5.64 & 0.12-0.21 \\
\end{tabular}
\end{ruledtabular}
\end{center}
\end{table}
%-------------------------------------------------------------------
The results from Model~I are consistent with those in Model~III,
which show that the bare mass of the $D_{s0}^*$ can be pulled down
significantly to the region close to the mass of the
$D_{s0}^*(2317)$ state. If we choose a bare mass from another quark
model, the obtained mass can even be consistent with the
experimental value.
In Model~I, a constant coupling is taken which would violate
Goldstone's theorem because the $\pi,~K$ and $\eta$ are Goldstone
bosons. Model~II has a large $\mu$ dependence, which makes it not
preferable for a phenomenological analysis. The Lagrangian for
Model~III is constructed from chiral symmetry, and the $\mu$
dependence is really small (for instance, the resulting mass of the
$D_{s0}^*$ would change to 2.37-2.41~GeV if we choose $m_c=1.35$~GeV
as the subtraction point, and the change is no more than 10 MeV).
Therefore, we choose the results in Model~III to give further
predictions. The absolute mass shifts in Model~III are listed in the
last column in Table~\ref{tab:results}. The  mass of the $D_0^*$ is
consistent with the experimental data
$2352\pm50$~MeV~\cite{Yao:2006px}. One strong decay channel $D\pi$
is open for the state, and the decay width of this channel should
give the dominant contribution to the width of the $D_0^*$. Then the
width of the $D_0^*$ can be obtained from
\begin{equation}
\Gamma_{D_0^*} = {h'^2\over2f_{\pi}^2M_{D_{0}^*}} {3\over2}{\rm
Im}G^{\rm III}_{D\pi}(M_{D_{0}^*}^2).
\end{equation}
Using the mass $M_{D_{0}^*}$ from Table~\ref{tab:results}, we
obtain
\begin{equation}
\Gamma_{D_0^*} = \text{ 99 -- 167 MeV}~.
\end{equation}
The result is roughly consistent with the experimental width for the
$D_0^*$, which was reported as $276\pm21\pm63$~MeV by the Belle
Collaboration~\cite{Abe:2003zm} and $240\pm55\pm59$~MeV by the FOCUS
Collaboration~\cite{Link:2003bd}. Both the mass and the width
suggest the observed $D_0^*$ state can be the dressed $c{\bar q}$ state.
Note that in Ref.~\cite{Guo:2006fu}, two molecular states were
predicted, and both of them are not consistent with the data.
Certainly, the uncertainty of the data is large so far, and more
precise measurements are highly desirable.
Another noticeable result is that the mass shifts in the bottom
sector are about twice of those in the charm sector, as can be seen
from the last column in Table~\ref{tab:results}. Based on heavy
quark symmetry, and assuming the $D_{s0}^*(2317)$ is a $DK$ bound
state, the mass of the $B\bar K$ bound state was predicted as
5733~MeV~\cite{Rosner:2006vc} which was confirmed by a dynamical
calculation~\cite{Guo:2006fu}, larger than the mass region obtained
here by dressing the $b{\bar q}$ state. That means if a $B_{s0}^*$
state with a mass which is much smaller than 5733~MeV were found, it
would probably be a dressed $b\bar q$ state rather than a $B\bar K$
bound state, or the $D_{s0}^*(2317)$ would not be a $DK$ bound
state. Similar to that of the $D_0^*$, the width of the $B_0^*$ can
be estimated as
\begin{equation}
\Gamma_{B_0^*} = \text{ 62 -- 100 MeV}~.
\end{equation}

\section{Summary}

The $D_{s0}^*(2317)$ is considered as a $DK$ molecular state by many
authors because in the quark model its mass simply comes out too
high. However, hadronic loops can pull down its mass. To make the
calculation of such an effect quantitative, we assume that the
hadronic-loop induced mass shifts of a hadron vanish at some point.
The point is chosen as $\sqrt{s}=m_D$ for the charm sector and $m_B$
for the bottom sector. Then we calculate the mass shifts of the
heavy scalar mesons by using three different types of coupling. The
input bare masses are taken from the Godfrey-Isgur quark
model~\cite{godfrey}. The mass of the $D_{s0}^*$ state is lowered
significantly, and it can even be pulled down to 2317 MeV if we
adjust the bare mass in the predicted region from different quark
models. That means the simple argument against the quark model for
too high masses is not valid. What the quark model predicts is just
the bare masses of hadrons. In order to compare with the
experimental spectroscopy, the bare hadron masses need to be
dressed~\cite{Capstick:2007tv}. Note, however, that such a dressing
is always model-dependent and must be considered in different
approaches, as done here.
Because chiral symmetry is fulfilled and the $\mu$ dependence is
small, we choose Model~III to give further predictions. The results
of both mass and width show that the observed $D_{0}^*$ could be a
dressed $c\bar q$ state. We also give predictions for the dressed
bottom scalar mesons. The mass of the $B_{s0}^*$ is smaller than
that of the $B\bar K$ bound state which was obtained assuming the
$D_{s0}^*(2317)$ be a $DK$ bound state~\cite{Rosner:2006vc}. Precise
experimental data from $B$ factories are highly desirable to test
the dressing mechanism.
We would like to stress that a consistent treatment of the mass and
width is required. The difference from the molecular state will
presumably be revealed in the decay pattern with into various
channels, this is a further step to be investigated.

\begin{acknowledgments}
We would like to thank Ch. Hanhart for useful discussions. The work
is partially supported by the Helmholtz Association through funds
provided to the virtual institute ``Spin and strong QCD''(VH-VI-231)
and by the EU Integrated Infrastructure Initiative Hadron Physics
Project under contract number RII3-CT-2004-506078.

\end{acknowledgments}

\end{document}